\documentclass[journal=aesccq,manuscript=comment]{achemso}

\makeatletter
\let\l@addto@macro\relax
\def\acs@author@fnsymbol#1{}
\makeatother
\usepackage[fontsize=11pt]{scrextend}

\usepackage[utf8]{inputenc}
\usepackage[version=4]{mhchem} 
\usepackage{csquotes}
\usepackage{wasysym}
\usepackage{hyperref, doi}
\usepackage{siunitx}
\usepackage[sort&compress,numbers,super]{natbib}
\setkeys{acs}{doi = true} 

\usepackage{tikz}
\usetikzlibrary{shapes}
\newcommand{\markercircle}{\raisebox{0.7pt}{\tikz{\node[draw,scale=0.3,circle,fill=black!100!](){};}}}
\newcommand{\markertriangle}{\raisebox{0.7pt}{\tikz{\node[draw,scale=0.3,regular polygon, regular polygon sides=3,fill=none](){};}}}

\SectionNumbersOn

\author{Olivier Durif}
\email{olivier@durif.fr}
\date{September 2023}

\title[Commentary Mortada 2022]{Commentary Regarding the CRESU-SIS Experiment: Concerns About the Uniform Supersonic Flow Reactor}

\begin{document}




\begin{abstract}
This commentary addresses the anomalies in the results reported from the CRESU-SIS experiment at the Institute of Physics of Rennes, France. This experimental setup is dedicated to studying ion-molecule kinetic in the gas phase at very low temperatures using a uniform supersonic flow reactor. A reinterpretation of the latest study performed with this instrument highly suggests a dramatic decrease in flow density upon the injection of neutral reactants. 
In particular, these concerns can be related to the diffusion effects prevalent in the reported results on the vast majority of the kinetics experiments conducted with a uniform supersonic flow reactor.
The scientific community in the field of low-temperature chemical kinetics in uniform supersonic flow would greatly benefit from being aware of and comprehending these highlighted anomalies because the evidence in this commentary calls into question many of the results published to date.
\end{abstract}

\section{Introduction}

\quad The first uniform supersonic flow (USF) reactor for kinetic studies between ions and neutrals at very low temperatures was built in 1984 in an aerodynamics laboratory by Rowe and collaborators\cite{rowe_study_1984}. The ion production was achieved using an \SI{20}{\kilo\electronvolt} electron gun positioned transversely to the flow at the nozzle exit. This approach ionized both the carrier gas and reactants indiscriminately, producing ion fragments and hot electrons, which made it difficult to control the chemical species involved in reactions. To address this problem, in 1989, the pioneers of these studies developed a device that generated ions outside the supersonic expansion and selected them using a quadrupole before injecting them directly into the nozzle via a small opening in the nozzle wall\cite{rowe_mass-selected_1989}. This allowed the ions to be relaxed along with the neutrals in the diverging section of the nozzle, where the density and the number of collisions are higher than at the nozzle exit. In this pioneering 1989 publication, the authors also discuss problems related to ionization using the electron gun, as well as two other opened possibilities to overcome their limitations: injecting ions into the reservoir or at the nozzle exit. Explicitly, they state, 
\begin{displayquote}
\textit{``it is evident that to overcome these problems the electron beam should be replaced by ‘something’ comprising an ion source and a quadrupole mass filter as in the SIFT injector. This ‘something’ cannot be incorporated in the nozzle chamber where the neutral reactant is injected: the reaction would occur prior to expansion so that the primary ion count rate would not be high enough at the nozzle exit for proper measurements, neither can it penetrate into the core of the flow without seriously damaging the supersonic flow uniformity''.}
\end{displayquote}

After the closure of the Meudon SR3 wind tunnel in the early 1990s, Rowe established a laboratory focused on experimental astrophysics in Rennes and, together with his colleagues, rebuilt the supersonic flow reactor for ion-neutral studies, using an electron gun positioned at the nozzle exit, as originally designed\cite{rowe_falp_1995}. Subsequently, other researchers have continued to work with USF reactors, but because they were no longer attached to an aerodynamics laboratory, some key expertise relaying to supersonic flows has ineluctably waned, such as the design of de Laval nozzles. This is illustrated, by the incorrect statement in\cite{joalland_mass-selective_2019} that \textit{``the technique relies on de Laval nozzles designed by solving the nonlinear Navier–Stokes equations for viscous fluids''}. As is commonly known, the Navier-Stokes equations describe the motion of Newtonian fluids, so they can be used to compute the flow profiles in a nozzle, but certainly not the shape of the nozzle itself. The calculation of de Laval nozzle design is generally performed by solving a potential equation for compressible flow, approximated by geometric considerations, and ultimately computed using the method of characteristics\cite{owen1952improved}. Nevertheless, once the geometrical shape of the de Laval nozzle is defined, a supersonic flow through this object can be simulated with computational fluid dynamics tools, which then rely on solving Navier-Stokes equations. 

That being said, in 2019, the CRESU-SIS instrument was modified by adding an ion selector at the nozzle exit\cite{joalland_mass-selective_2019}. However, in this work, the warning issued 30 years earlier by the pioneers\cite{rowe_mass-selected_1989} that penetrating into the core of the flow could seriously damage the uniformity of the supersonic flow was disregarded.

\section{Anomalies in the CRESU-SIS data}

\quad Regrettably, despite being funded by the French National Research Agency (ANR) and the French national research programs supported by CNRS, CNES, and CEA, the authors did not grant access to the data upon request. Consequently, in the following, we are limited to analyzing the most recent study carried out using CRESU-SIS, based solely on the kinetics data presented in Figure 3 of their latest article\cite{mortada_kinetics_2022}. Nevertheless, the same analysis is transposable to their previous study \cite{joalland_mass-selective_2019} as well.

\subsection{Pseudo first-order kinetic}

\quad In this latest work performed with CRESU-SIS\cite{mortada_kinetics_2022}, the authors studied two distinct reactions. The first reaction concerns the ion \ce{N2+} with propyne (\ce{CH2CCH2}). The second reaction concerns the ion \ce{N2+} with allene (\ce{CH3CCH}). Both reactions were studied under the same conditions, but separately.

The reaction investigated is the charge transfer reaction
\begin{center}
 \ce{N2+ + C3H4 ->[k] N2 + C3H4+},
\end{center}
where $k$ is the bimolecular rate coefficient of the reaction. Assuming a pseudo first-order reaction ($\ce{[C3H4]} >> \ce{[N2+]}$), the kinetic is expressed depending on the \ce{N2+} ion
 \begin{equation}
  \frac{\mathrm{d} \ce{[N2+]}}{\mathrm{d}t} = -k^\prime \ce{[N2+]}
  \label{eq:cinetique}
 \end{equation}
with, $k^\prime $ is the pseudo first-order rate coefficient such that
\begin{equation}
    k^\prime = k \ce{[C3H4]}.
    \label{eq:eq_k}
\end{equation}
To determine the bimolecular rate coefficient, $k$, equation \eqref{eq:cinetique} can be easily integrated
 \begin{equation}
  \ce{[N2+](t)} =  \ce{[N2+]_0} \,\, e^{{- k^\prime (t-t_0)}}.
 \end{equation}

In practice, the authors of the study varied the concentration of \ce{[C3H4]} and recorded the kinetics based on the signal of \ce{N2+} along the central axis of the USF. In this study, $t_0 = 0$ is where the ions merge the core of the flow. To determine the bimolecular rate coefficient of the reaction $k$, the authors of the study then decided to perform a two-parameter regression, $k^\prime$ and $\ln{\ce{[N2+]_0}}$, in a semi-logarithmic space, for each kinetics at each concentration, such that
 \begin{equation}
  \ln \ce{[N2+]} =  \ln{\ce{[N2+]_0}} - {k^\prime t}
  \label{eq:semi_log}
 \end{equation}
is an affine function. Once $k^\prime$ has been computed, the value of $k$ is deduced using Eq.~\eqref{eq:eq_k}. This principle is straightforward and forms the basis for many kinetic studies.

\subsection{Regression on the kinetics data}

\begin{figure}[h]
    \centering
    \includegraphics[width=\textwidth]{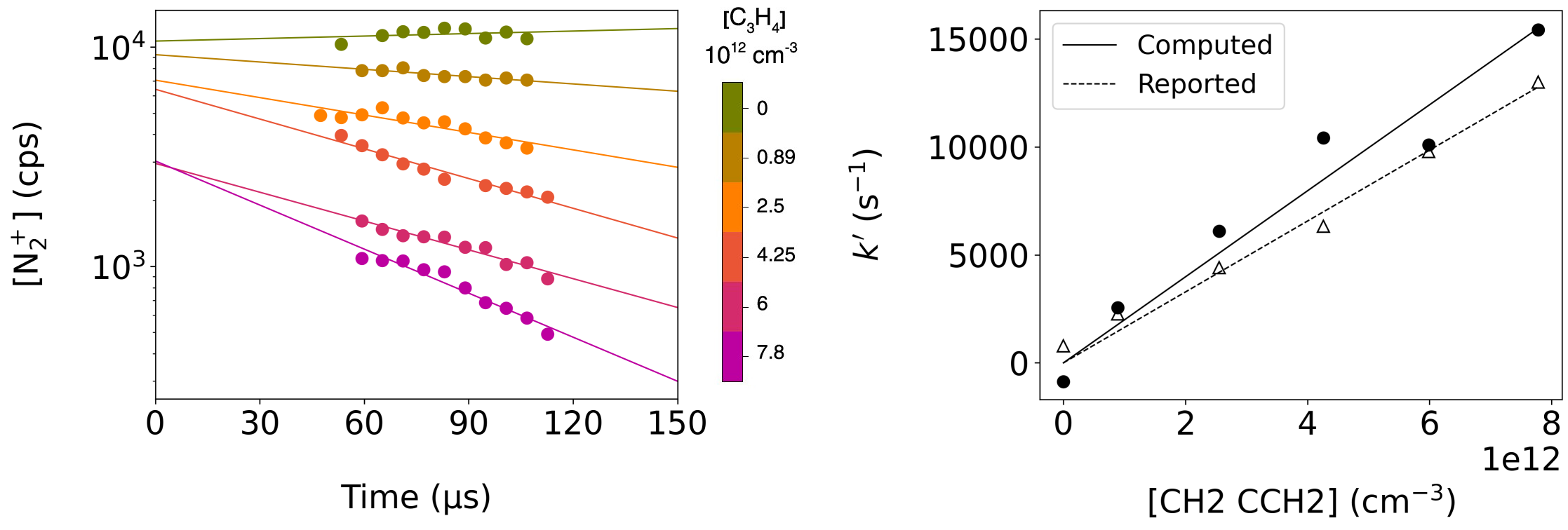}
    \caption{On the left figure, kinetics traces are fitted with Eq.~\eqref{eq:semi_log} to compute the pseudo first-order coefficient $k^\prime$ which is reported for each curve as a point on the right figure. On the right figure, $k^\prime$ is thus fitted with Eq.~\eqref{eq:eq_k} to compute the bimolecular rate coefficient $k$. Here $k^\prime$ computed by ourselves (\protect\markercircle) is compared to the one reported by the authors\cite{mortada_kinetics_2022} (\protect\markertriangle), yet using the same dataset.}
    \label{fig:cinetique}
\end{figure}

\quad Let us take the kinetics data from Figure 3 of this latest work with CRESU-SIS\cite{mortada_kinetics_2022} about reaction \ce{N2+ + CH2CCH2} at \SI{24}{\kelvin} and perform our own linear regression on the dataset (Fig. \ref{fig:cinetique}, left). The value of the bimolecular rate coefficient $k$ reported by the authors (Fig. \ref{fig:cinetique} right, \markertriangle~symbol) and the value one may calculate (Fig. \ref{fig:cinetique} right, \markercircle~symbol) differs significantly. To be clear, the results of the pseudo first-order coefficients reported by the authors do not correspond to the kinetic dataset they used. Among them, one value is shifted by almost a factor of two from the expectation deriving from the kinetic traces data. Finally, the value of $k = \SI{2.0e-9}{\centi\meter\cubed\per\second}$ is obtained by our own linear regressions on the dataset, using the ordinary least square method. This result contrasts with the value \SI{1.52 \pm 0.16 e-9}{\centi\meter\cubed\per\second} reported by the authors. This latest value is surprisingly found to be shifted from the experimental data but in perfect agreement with the predicted theoretical value. 
Indeed, the authors assert and repeat that their value is \textit{``in good agreement with the Langevin value of $k_L = \SI{1.38 e-9}{\centi\meter\cubed\per\second}$''} and further state that \textit{``the experimental values obtained in this work and the estimated values obtained from the simple capture models of Langevin and Su and Chesnavich, illustrate their remarkable agreement.''}. 

Since discrepancies are observed, it is crucial for the authors to explain them and take necessary measures to prevent such errors from occurring among the seven other bimolecular rate coefficients reported in this study\cite{mortada_kinetics_2022}. 
The authors of this work are encouraged to openly provide their data for scrutiny. This would allow anyone to verify the authenticity of their results.

\subsection{Main concern about the ions initial concentration variability}

\quad In the study\cite{mortada_kinetics_2022}, the authors kept the initial concentration of \ce{[N2+]} constant while varying the concentration of \ce{[C3H4]}. One should take a close look at the Figure 3 in\cite{mortada_kinetics_2022} (Fig.~\ref{fig:cinetique} here) and ask this question, \emph{why is the initial concentration of \ce{[N2+]_0} inversely proportional to the \ce{[C3H4]} rather than constant?}

\begin{figure}[H]
    \centering
    \includegraphics[width=\textwidth]{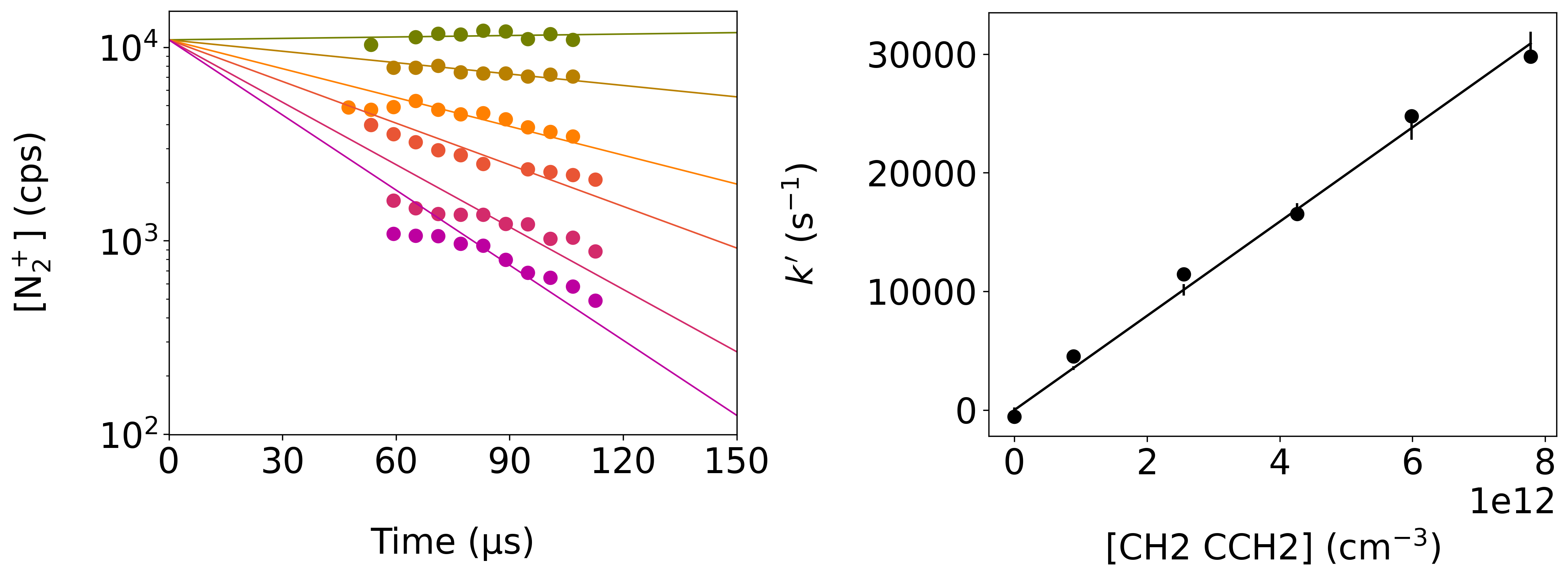}
    \caption{Regression by fixing \ce{[N2+]_0} constant. The $k$ obtained is \SI{3.97e-09}{\centi\meter\cubed\per\second}, but the value is not reliable since the model doesn't fit with the kinetics data. The error bars on the right figure correspond to the standard deviation obtained by the regressions on the right figures.}
    \label{fig:cinetique_1param}
\end{figure}

The regressions carried out below to replicate the linear fit of the authors (see Fig.~\ref{fig:cinetique}) are two parameters, with $k^\prime$ the slope of the curve and $\ln{\ce{[N2+]_0}}$ the intercept (see Eq.~\eqref{eq:semi_log}). However, $\ce{[N2+]_0}$ is a constant experimental parameter for each kinetics, thus it is better to constrain the regression on the kinetics traces to a single parameter, $k^\prime$. Let's set $\ln{\ce{[N2+]_0}} = 9.3$, corresponding to approximately 11000 counts per second for $\ce{[N2+]_0}$ ($\exp 9.3 =10938$) in order to perform a single parameter fit (Fig.~\ref{fig:cinetique_1param}).
The kinetic model does not match with the kinetics data (Fig.~\ref{fig:cinetique_1param}, left) although the regression computed on the bimolecular coefficient is excellent (Fig.~\ref{fig:cinetique_1param}, right) and the standard deviation very low (as reported by the error bars on Fig.~\ref{fig:cinetique_1param}, right). Here, the author's own results actually highlight a loss rate of $\ce{[N2+]}$ proportional to $\ce{[C3H4]}$. This loss rate is also reflected in the value of $\ce{[N2+]_0}$, so strongly suggesting that it is not due to the studied reaction, but rather to an experimental artifact.

Two possible interpretations of these results have been identified, which are discussed in more detail in the following sections:
\begin{itemize}
\item Either the $\ce{[N2+]_0}$ is not constant because an amount reacts with the \ce{[C3H4]} prior to the studied reaction. For instance, ions could react into the boundary layers of the flow, before the \ce{N2+} enters the uniform cold kernel.
\item Either the entire $\ce{[N2+]}$ decreases due to an experimental bias. This suggests flow behavior that exhibits non-uniformity in density (and so temperature as well). In such case, the $\ce{[N2+]_0}$ could fall due to damage on the flow uniformity upon introduction of hydrocarbons into the carrier gas.
\end{itemize}

\subsubsection{Hypothesis of a boundary layers phenomenon}

\quad A reaction between $\ce{N2+}$ and $\ce{C3H4}$ in the boundary layers certainly occurs, but seems highly unlikely to explain the reported data. Indeed, the transit time of the $\ce{N2+}$ ions in the boundary layers is extremely brief and the density in the boundary layers is lower than in the isentropic core of the flow. However, an accurate computation would be necessary to quantify this phenomenon. 

To provide some ideas, the deflector used to merge the ions with the supersonic flow brings approximately \SI{10}{\electronvolt} to the ions. It means that the ions merge with the adiabatic expansion at a collisional temperature of around \SI{100000}{\kelvin}. This very high collisional temperature could potentially lead to particularly high reactivity in the boundary layers compared to the isentropic core. However, here is another problem. For a single ion, the energy provided by the electrostatics fields is considerably larger than the thermal energy of the flow (which is only a few tens of \SI{}{\kelvin} thus a few \SI{}{\milli\electronvolt}). This could play a role, especially since the electrostatic field of the deflector probably leaks into the flow, as does the polarized nose of the spectrometer. However, the authors of these studies could rightfully argue that their estimated ion density is on the order of \SI{e5}{\cm\cubed}, while the density of neutral gas is on the order of \SI{e16}{\cm\cubed}, and therefore, even if the ions were hot, they would quickly relax by collisions in the flow and that the energy brought by the ions is negligible. It is true that the energy brought by the ions is overall low, even if the ion density could be seriously underestimated. In fact, the mean free path under these pressure and temperature conditions in the flow is on the order of \SI{10}{\micro\meter}. With an electrostatic field of \SI{1}{\volt\per\centi\meter}, an ion would acquire only sub-\SI{}{\micro\electronvolt} energy between two collisions, which is indeed negligible. Therefore, it is quite unlikely that the thermal energy provided by the ions is sufficient alone to heat so much the uniform supersonic flow.

\subsubsection{Hypothesis of a non-uniform flow}


\quad The suggested possibility of a decreasing density of the supersonic flow along its center axis upon the gas mixture is an avenue that the scientific community in the field of kinetic studies in USF should rigorously explore. This topic is deeply complex, but such a hypothesis provides a successful explanation of these data as well as an alternative interpretation to a number of other results (see Section~\ref{sec:other_results}). 

As improbable as it may seem, this issue of rarefied supersonic flow diagnosis under real gas conditions is poorly established and literature on the subject is sorely lacking. Actually, the classical flow diagnosis measurement is performed using a Pitot tube but this method has been found to be not reliable in real gas conditions, for rarefied and supersonic flow\cite{durif_nouvel_2019}. Indeed this classical method consists of an indirect measurement based on the impact pressure of the shock wave attached to a probe and on the Rayleigh Pitot tube formula which has been found ineffective experimentally in real gas conditions when accounting for the gas mixture as well as compressibility errors becomes crucial and makes the Rayleigh Pitot tube equations inappropriate\cite{durif_nouvel_2019}. 

As an alternative to Pitot measurement, the investigators could simply monitor the secondary vacuum pressure in the spectrometer chamber while varying the spectrometer-nozzle distance to confirm or refute this hypothesis of flow uniformity breakdown. This easily accessible information, obtained from experiments where the flow is sampled into a secondary vacuum chamber, directly addresses concerns about flow uniformity. 

More broadly, directly monitoring vacuum pressure using a Pirani gauge through a skimmer appears to be an ideal alternative to Pitot measurements for the diagnosis of the rarefied supersonic flow of a complex gas mixture\cite{durif_nouvel_2019}. 

Obtaining information about flow uniformity is crucial as the credibility of uniform supersonic flow kinetics relies on it. Especially under real gas conditions, regarding the limitations of Pitot probe measurements this issue deserves special attention and specific in-depth investigation would be required. However, even this information about the flow density could not be sufficient to ensure the relevance of kinetic results if the flow remains dense yet inhomogeneous\cite{zarvin_features_2017}. 



\subsection{Reported branching ratio}

\quad The branching ratio reported in this study also supports the hypothesis of non-uniform flow. Indeed, when the authors of this work increased the allene concentration, they observed a branching ratio that evolved in a manner consistent with the temperature (see Figure 5 and Figure 6a)\cite{mortada_kinetics_2022}. Normally, for a given pressure and temperature conditions, the branching ratio should be independent of the allene concentration. However, since the observed branching ratio changed with increasing allene concentration, it suggests that the flow warm-up with the allene. The interpretation of a gas mixture effect on flow uniformity provides a satisfactory explanation for this observed result. 

\subsection{Kinetic trace following the product}

\quad When possible, the kinetics traces should be recorded according to the growth of the product rather than the decay of the reactants. Especially in the CRESU-SIS experiment, the detection is performed using a quadrupole and a channeltron, so the authors could have swept away any doubt about their kinetics traces by tracking the signal of the product. This elementary precaution would have allowed to avoid artifacts related to a density drop when following the reactant.
Notably, studies investigating chemical kinetics in USF, with a focus on tracking the products, are extremely rare, if none, because most of the measurements conducted thus far have predominantly concentrated on the detection of radicals using fluorescence techniques.

\section{Others kinetics experiments in USF}\label{sec:other_results}

\quad The consideration of the non-homogeneity of supersonic flow provides a new perspective that lead to the reinterpretation of numerous chemical kinetics results published using a de Laval nozzle reactor.


\subsection{Mass spectrometry experiments}


\quad Various experiments with CRESU-SIS have reported bell-shaped curves symptomatic of increasing-decreasing trends when following the products\cite{biennier_low_2014,bourgalais_elusive_2016,joalland_low-temperature_2016}. The kinetics of clusters in USF reported the same trends too\cite{bourgalais_low_2016,bourgalais_propane_2019}. We suggest that such a bell-shaped curve when following the products is strongly indicative of a density fall. The kinetic trace of the carrier gas should be able to confirm or refute this hypothesis, but unfortunately, this information has not been reported in the scientific literature. Therefore, from the sole reading of the articles, we are unable to determine whether the studies were conducted under uniform flow conditions or not. 

Meanwhile, others have observed in similar conditions an explosion in the size of the clusters beyond a threshold \cite{lippe_water_2018}. This observation could be also consistent with a sudden increase in collisions due to a transition to a turbulent flow regime. This physical phenomenon could simply explain the mysterious bottleneck reported in the clustering processes\cite{bourgalais_low_2016,lippe_water_2018}. 

One can also observe mass spectrometry studies that do not present an intercept for the pseudo first-order kinetic coefficient passing through zero \cite{lee_direct_2000,soorkia_reaction_2010,soorkia_airfoil_2011}, contrary to the requirement of Eq.~\eqref{eq:eq_k}. Unfortunately, this point is not often discussed in mass spectrometry experiments, although it has been attributed to some loss processes\cite{bouwman_bimolecular_2012} or secondary reactions\cite{bouwman_reaction_2013}, such behavior is again consistent with a non-uniformity of the flow.

This possibility of breaking the uniformity of supersonic flow in the presence of reactants to explain these kinetics trends was discussed about the dimerization of formic acid with CRESUSOL\cite{durif_nouvel_2019} but for which only a very little literature exists on this topic. 

\subsection{Fluorescence experiments}

\quad The scientific literature about chemical kinetics in USF reactors primarily consists of experiments focused on radical-neutral reactions using PLP-LIF techniques. Most of these experiments typically rely on monitoring the decay of a fluorescent radical to track the kinetics because the formation of a stable product cannot be detected. In such an approach, a density fall of the flow appears as a decreasing signal and superposes to the decreasing signal expected because of a chemical reaction when following the reactant. This density fall could thus be mistakenly confused as an increase in reaction efficiency. 

While most fluorescence studies in USF are conducted at a fixed distance between the nozzle and the detector, the density at the probe point can fluctuate depending on the reactant concentration. This effect is especially pronounced if the distance between the nozzle and the detector is too long. Therefore, some studies may be operated in a transient regime where the flow is not homogeneous and where a small perturbation can alter the flow behavior.

To eliminate any artifact resulting from a misinterpretation of a density fall of the flow, some simple criteria can be checked:
\begin{itemize}
    \item The pseudo first-order kinetic coefficient has to fit originally at zero without co-reagent and when no secondary reaction is competing (see Eq.~\eqref{eq:eq_k}).
    \item The maximum amplitude (i.e. initial concentration) of the reacting radical produced by the pump laser must remain constant regardless of the co-reagent.
    \item The signal has to be identical over the same period of time when changing the point of the probe in the flow and reducing the distance between the nozzle exit and the detector.
\end{itemize}

These criteria are elementary, yet, the latest review on this topic\cite{cooke_experimental_2019} enabled to quickly compile a list of studies that fail to satisfy the first criterion, because the pseudo first-order coefficient does not cross zero, and do not even address why\cite{sims_ultra-low_1994,bocherel_ultralow-temperature_1996,le_picard_experimental_1997,brownsword_kinetics_1997,jaramillo_consensus_2002,vakhtin_low-temperature_2003_2,goulay_reaction_2006,goulay_low-temperature_2006,paramo_experimental_2006,vohringer-martinez_water_2007,hansmann_kinetics_2007,sabbah_understanding_2007,paramo_rate_2008,berteloite_low_2008,liessmann_primary_2009,berteloite_low_2010,morales_experimental_2010,trevitt_reactions_2010,bennett_chemical_2010,berteloite_kinetics_2010,liessmann_reaction_2011,morales_crossed_2011,daranlot_gas-phase_2012,hickson_unusual_2013,shannon_fast_2014,shannon_combined_2014,stubbing_gas-phase_2015,sleiman_low_2018,sleiman_gas_2018,blazquez_experimental_2019,hickson_kinetic_2020}. 


Otherwise, in most of the publications where this unexpected non-zero intercept is discussed, it is often attributed to diffusion loss\cite{herbert_rate_1996,james_rate_1997,vakhtin_low-temperature_2001,vakhtin_low-temperature_2003,spangenberg_low-temperature_2004,daugey_kinetic_2005,bergeat_low_2009,shannon_observation_2010,daranlot_gas-phase_2010,daranlot_elemental_2012,cheikh_sid_ely_low_2013,shannonAcceleratedChemistryReaction2013,gomez_martin_low_2014,hickson_c3p_2015,hickson_ring-polymer_2015,caravan_measurements_2015,hickson_experimental_2016_2,hickson_quantum_2016,nunez-reyes_kinetic_2017,douglas_low_2018,douglas_low_2019,westMeasurementsLowTemperature2019,gupta_low_2019,messinger_rate_2020,blazquezGasphaseKineticsCH2020,hickson_experimental_2022,a.westExperimentalTheoreticalStudy2023,douglasExperimentalTheoreticalAstrochemical2023} and sometimes because of the impurities too. This idea of loss by diffusion is the commonly accepted interpretation in this field of research. But what \textit{diffusion} means physically except that the flow is not homogeneous?

Often, secondary reactions also occur and are accountable for this non-zero intercept \cite{sharkey_pressure_1994,herbert_rate_1996,chastaing_neutralneutral_1998,chastaing_neutralneutral_1999,lee_rate_2000,vakhtin_kinetics_2001,murphy_laboratory_2003,nizamov_rate_2004,nizamov_kinetics_2004,mullen_low_2005,goulay_low-temperature_2006,daugey_reaction_2008,berteloite_low_2010_2,berteloite_low_2011,daranlot_low_2013,nunez-reyes_low_2019,nunez-reyes_tunneling_2020,hickson_kinetic_2020,hickson_experimental_2022,vanuzzo_reaction_2022}. 
While secondary reactions occur, it is assumed that they do not affect the slope of the pseudo first-order curve. However, this assumption is dangerous because if the secondary reactions are significant enough to explain a non-negligible intercept, the hypothesis that this secondary reaction becomes negligible or does not affect the primary reaction looks questionable. In a general way, when considering both or multiple reactions that compete simultaneously, the analyses become often complex and debatable.

Furthermore, when following a reactive radical that is not in the ground state, the quenching of the radical is often  given as an explanation for this intercept\cite{goulay_reaction_2006,paramo_experimental_2006,vohringer-martinez_role_2010,loison_experimental_2014,meng_theoretical_2016,grondin_low_2016,hickson_experimental_2016_1,hickson_low-temperature_2017,lara_experimental_2017,nunez-reyes_combined_2018,nunez-reyes_experimental_2018,nunez-reyes_rate_2018,nunez-reyes_kinetics_2018,nunez-reyes_experimental_2019,hickson_kinetic_2022}.  One also notes that intrinsic fluorescence has also been put forward as an explanation\cite{sanchez-gonzalez_low-temperature_2014}. Actually, the idea of quenching or intrinsic fluorescence formally amounts to take into account one or multiple secondary reactions in competition. It may also be noticed that for the same radical, but depending on the system and the nozzle, the intercept varies considerably\cite{nunez-reyes_low_2018,nunez-reyes_low_2019}. Therefore, it is legitimate to wonder whether it is always the same process that is at work.

Generally, when product detection is possible, authors attempt bi-exponential fits to replicate the bell-shaped curves of a reaction product with the idea of accounting for secondary reactions and/or diffusion\cite{hickson_ring-polymer_2015,bourgalais_c3p_2015,hickson_c3p_2015,hickson_temperature_2016,hickson_experimental_2016_1,hickson_experimental_2016_2,nunez-reyes_kinetic_2017,nunez-reyes_kinetics_2018,hickson_experimental_2022}.  In such studies, it is particularly important to verify that the loss coefficient is constant from one kinetic to another, regardless of the co-reagent concentration, to ensure that the flow is well uniform.

Sometimes, it is difficult to make one's own judgment because information about the pseudo first-order coefficient is not reported\cite{geppert_comparison_2000,chastaing_rate_2000,chastaing_direct_2000,canosa_experimental_2007,daranlot_revealing_2011} even though diffusion may be discussed\cite{tizniti_low_2010}.

Sometimes, when the study is also conducted under varying flow pressure, extrapolations to zero pressure indicate that the reaction is still effective \cite{shannon_combined_2014,heard_rapid_2018}. The authors interpret the results as evidence of a quantum tunneling effect although the uniformity of the flow is not questioned.

Studies on catalysis in the presence of water\cite{vohringer-martinez_water_2007,vohringer-martinez_role_2010}, 
for example, become much clearer when one considers that the density of the flow precipitate due to the excess of water. In this same study\cite{vohringer-martinez_water_2007}, one can observe that the intercept of the pseudo first-order coefficient perfectly passes through zero at ambient temperature, but not under supersonic flow conditions.

It has also been noticed in the literature that the intercept depends on the laser alignment\cite{taylor_pulsed_2008,gupta_low_2019,thawoosKineticsCNReactions2023} which is perfectly consistent with the probe of a non-uniform flow.

In some experiments, a significant difference in the maximum amplitude of the exponential decay is observed between kinetics with and without reactants\cite{hickson_kinetic_2020,vanuzzo_reaction_2022}, or when increasing the co-reagent\cite{douglasGasphaseReactionNH2022}. This constitutes the second criterion proposed earlier because a drop in density due to the gas mixture provides a plausible explanation for this result. 

Furthermore, one might also observe studies that do not meet the first and second criteria\cite{hansmann_kinetics_2007}.

The breakdown of the flow uniformity could also explain the apparent kinetic slowdown observed in PLP-LIF measurements when increasing the reactant concentration, which has previously been attributed to secondary reactions with clusters, especially dimers\cite{sims_ultralow_1994,hamon_low_2000,jimenez_development_2015,antinolo_reactivity_2016,jimenez_first_2016,ocana_gas-phase_2019}. 

It is very important to note that some studies have reported results with such USF reactors where the intercept of the pseudo first-order curve passes through zero\cite{sims_ultralow_1994,leonori_crossed-beam_2009,hickson_low_2010,tizniti_rate_2014,jimenez_development_2015,antinolo_reactivity_2016,jimenez_first_2016,ocana_gas_2018,ocana_gas-phase_2019,gonzalez_gas-phase_2022}. Thus when everything suggests that the experiment is conducted under good uniformity conditions, it is proved that there is no intrinsic loss due to diffusion in USF.

On top of that, when authors have attempted to probe by varying the nozzle-detector distance in PLP-LIF\cite{canosa_rate_2001,picard_sipj_2001,picard_comparative_2002,picard_comparative_2002,geppert_rate_2004,le_picard_experimental_2004,canosa_experimental_2004}, they reported an intercept passing through zero. Experiments exhibiting a non-zero intercept because of diffusion losses could be confirmed by varying the probed length to ensure that the diffusion is not because of the too long nozzle-detector distance.

The measurement performed by varying the nozzle-detector distance could be utilized by researchers whose experiments show a non-zero intercept due to diffusion. In such a manner, they may confirm their results and demonstrate that the diffusion arises from the uniform supersonic flow at any length and is not an experimental artifact.

Also, probing the co-reagent could be helpful to address this controversy. The co-reactant which is assumed to be constant, can sometimes be detected\cite{douglasExperimentalTheoreticalAstrochemical2023}. The temporal profile of this compound along the supersonic flow would directly answer these inquiries. If the signal remains constant along the flow, the flow is uniform. But if the hypothesis of a non-uniform flow holds true, the signal would exhibit a decay over time. If a decaying signal is recorded, it is of the utmost interest to answer whether the decay would be constant, because of constant diffusion velocity whatever the concentration in the carrier gas, or proportional to the concentration of this co-reactant, because of a mixture of gases affecting flow uniformity.

It is also noteworthy that a non-zero intercept seems less observed in the most analogous experiments but in cold cells\cite{pedersen_laboratory_1993} like at room temperature\cite{choi_h_2004,bennett_chemical_2010,hickson_experimental_2016_2,amedro_kinetics_2019,sun_kinetics_2022,sun_rate_2022}. But it is true that one can also notice that in some experiments, a non-zero intercept has been reported in cryogenic cells\cite{sharkey_pressure_1994}, as well as in room-temperature setups\cite{bergeat_reaction_2001,gannon_kinetics_2008,amedro_kinetics_2019,amedro_kinetics_2020,vanuzzo_reaction_2022,sun_kinetics_2022,sun_rate_2022}, despite it is only a small intercept most of the time. At room temperature, bell-shaped curves probed on products have been also observed\cite{lgannon_experimental_2010}. But the diffusion coefficient is temperature dependent, which is consistent with the fact that it has been observed that diffusion losses increase with temperature\cite{blitz_effect_2005}. Thus, it is surprising when the intercept reported at low temperatures in USF appears to be higher than at room temperature\cite{nunez-reyes_kinetics_2018}.

Finally, existing data are probably already sufficient to settle this controversy. But it would be necessary to go into the data from the PLP-LIF in USF studies and to investigate them with the idea of doing this verification and to understand what is behind the diffusion. In particular, when there is a good reason for a non-zero intercept as several reactions compete, it is necessary to verify precisely if the maximum amplitude of the initial radical signal immediately after the excimer laser fire is always equal to whatever the co-reagent concentration, and if not, why.


If these verifications lead to confirm these anomalies, what would be measured would not be a bimolecular coefficient of the reaction, but a coefficient of fall of the density of the flow proportional to the gas mixture. 

\section{Conclusion}

\quad The kinetics rate coefficients reported in the latest CRESU-SIS studies can be considered as an artifact. At least, this is a reasonable standpoint that should prevail as long as no one provides an answer to the simple question \textit{why is the initial concentration of the ion reactant inversely proportional to the neutral reactant, instead of being constant?}

What is highlighted here, in the reading of these data, is that the decay of the signal of the reactant is proportional to the concentration of the co-reagent. Therefore, it is likely possible that the reactant is not reacting with the co-reagent, but rather as an alternative explanation that the decay of the signal betrayed the density fall of the flow. This alternative explanation is fulfilling for CRESU-SIS experiment because it gave a satisfactory reason for why the ion density prior to the reaction is inversely proportional to the co-reagent, instead of being constant.

This alternative interpretation is also satisfactorily applicable to most of the experiments in USF reactor. It is especially applicable for kinetic studies depicting decreasing trends when following the product and experiments that do not present an intercept originally passing at zero because of diffusion. This observation could express this same phenomenon of a density fall.

These fundamental concerns raise crucial questions about the understanding of the USF reactors and the reliability of most of the results published so far. The scientific community in this field of research must be fully aware of these stakes and we urge to explore much further. 

\begin{suppinfo}
The data that support the findings of this commentary are openly available in Zenodo at

\href{https://doi.org/10.5281/zenodo.7685812}{https://doi.org/10.5281/zenodo.7685812}, reference number \href{https://doi.org/10.5281/zenodo.7685812}{7685812}.
\end{suppinfo}

\bibliography{biblio}

\end{document}